\documentclass[         %
aps,                    
prl,                    
showpacs,               
nofootinbib,            
twocolumn,              %
floatfix]               
{revtex4}               

\usepackage{amssymb}
\usepackage{graphicx}
\usepackage{amsmath}
\usepackage[hypertex]{hyperref}
\usepackage{amsfonts}

\begin{document}

\title{DNA Replication via Entanglement Swapping}
\author{Onur Pusuluk $^{1}$}
\author{Cemsinan Deliduman $^{2,3}$}
\affiliation{$^1$ \.{I}stanbul Technical University, Faculty of
Science and Letters, Physics Department, Maslak 34469, \.{I}stanbul,
Turkey} \affiliation{$^2$ Mimar Sinan Fine Arts University,
Department of Physics, Be\c{s}ikta\c{s} 34349, \.{I}stanbul, Turkey}
\affiliation{$^3$ Feza G\"{u}rsey Institute, \c{C}engelk\"{o}y
34684, \.{I}stanbul, Turkey}

\begin{abstract}
Quantum effects are mainly used for the determination of molecular
shapes in molecular biology, but quantum information theory may be a
more useful tool to understand the physics of life. Organic
molecules and quantum circuits/protocols can be considered as
hardware and software of living systems that are co-optimized during
evolution. We try to model DNA replication in this sense as a
multi-body entanglement swapping with a reliable qubit
representation of the nucleotides. In our model molecular
recognition of a nucleotide triggers an intrabase entanglement
corresponding to a superposition state of different tautomer forms.
Then, base pairing occurs by swapping intrabase entanglements with
interbase entanglements. We examine possible realizations of quantum
circuits to be used to obtain intrabase entanglement and swapping
protocols to be employed to obtain interbase entanglement. Finally,
we discuss possible ways for computational and experimental
verification of the model.


\end{abstract}


\maketitle

According to the central dogma of molecular biology, genetic
information stored in DNA is duplicated by replication and is used
by successive transcription and translation. During replication,
enzyme DNA polymerase (DNA\emph{pol}) recognizes the nucleotide
bases N = \{A, T, G, C\} of template DNA strand and finds their
complementaries \{\={A}=T, \={T}=A, \={G}=C, \={C}=G\} from the
surrounding environment for base pairings. Recognition interaction
between this single-stranded DNA (ssDNA) and polymerase enzyme is
one of the several unknown aspects of replication. Also, mechanism
used for finding the correct nucleotide from the surrounding
environment is a mystery. Since a lot of amino acids exist in the
active site of DNA\emph{pol} \cite{KE-1}, both experiments and
quantum chemical calculations are insufficient to clarify these
mysteries. However, there are some quantum information processing
models proposed for replication mechanism.

\begin{figure}[h]
\centering
\includegraphics[width=0.33\textwidth]{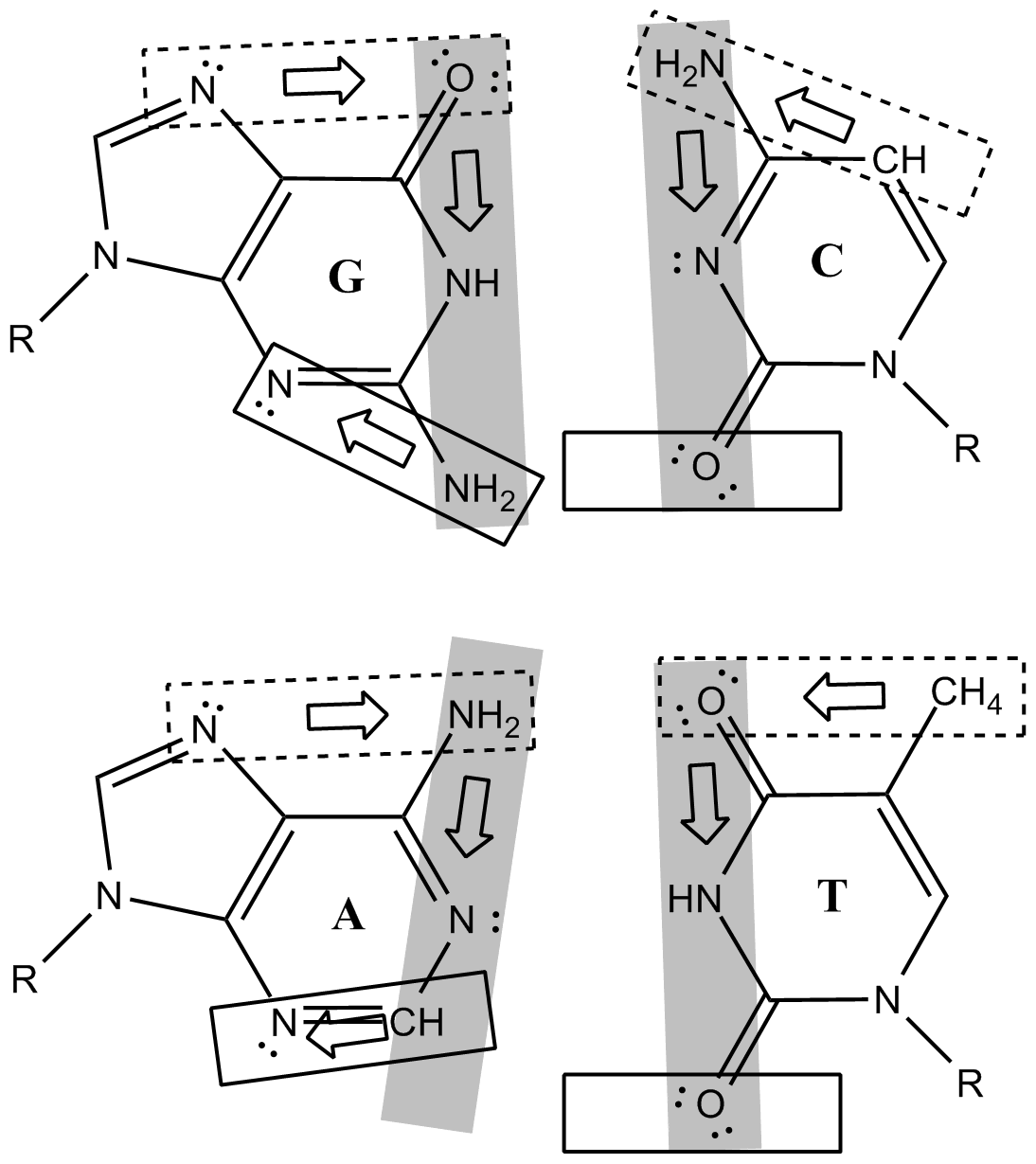}
\caption{Parts of the nucleotides: atoms can be grouped according to
region they will be found in DNA. Hoogsteen ($\leftrightarrow$major
groove), Watson-Crick ($\leftrightarrow$pairing plane), and Sugar
($\leftrightarrow$minor grove) edges are indicated respectively by
dashed, filled, and plain boxes. Arrows inside the boxes show the
order of the qubits used in qubit representation.} \label{F-parts}
\end{figure}

In 2001, Patel \cite{KE-2} formulated nucleotide selection from
surrounding environment as an unsorted database search and examined
the possibility of the use of Grover's algorithm \cite{KE-7} in the
evolutionary context. Although he achieved to model base pairing as
oracle in the algorithm, initiation of the algorithm requires the
superposition of four nucleotides which is not quite possible. In
the wave analogue of the algorithm \cite{KE-3} replication should
begin with the interaction of DNA\emph{pol} and all of four
nucleotides, but it is known that DNA\emph{pol} first binds to DNA
template in this process \cite{KE-1}. Recently, Cooper \cite{KE-4}
showed that molecular genetic transcription data of bacteriophage T4
is compatible with a quantum treatment in which enzyme makes a
measurement on coherent protons that causes an entanglement between
enzyme and protons. According to this model, enzyme-nucleotide
interaction takes place on the protons and electron lone pairs that
contribute to the base pairing. However, Watson-Crick ($WC$) edge of
the nucleotides (Figure \ref{F-parts}) can not be represented as
four orthogonal states by such a treatment. In this paper, we try to
model DNA replication as a multiparticle entanglement swapping
\cite{KE-6} with a reliable qubit representation of nucleotides. In
the model, molecular recognition of a nucleotide takes place on the
protons and electron lone pairs that are present on the Hoogsteen
($H$) edge.

Meanwhile, covalent structure of the nucleotide bases is not static.
Random, reversible and infrequent proton-coupled electron
delocalizations convert bases into rare tautomer forms (N$^{*}$,
G$^{\sharp}$, C$^{\sharp}$). In 1963, L\"{o}wdin \cite{HBPT-1}
claimed that tautomeric shifts on both paired DNA bases by a double
proton tunneling through hydrogen bonds can be responsible for
mutations. After some correlated ab inito calculations \cite{HBPT-2,
HBPT-3} that support L\"{o}wdin's claim, Villani \cite{HBPT-6,
HBPT-9} calculated more reliable potential energy surfaces by
density functional theory (DFT) method and found possible
A$\cdot$T$\rightarrow$A$^{*}$$\cdot$T$^{*}$,
G$\cdot$C$\rightarrow$G$^{\sharp}$$\cdot$C$^{\sharp}$ and
G$\cdot$C$\rightarrow$G$^{*}$$\cdot$C$^{*}$ transitions by concerted
or two step double proton-coupled electron transfers. Since
calculated transition probabilities are pure quantum mechanical
ones, tautomerization in double-stranded DNA should have a quantum
mechanical nature. However, cellular environment may cause a
decoherence effect on nucleotide base in both free nucleotide and
ssDNA cases. Therefore, we assume that enzyme-nucleotide base
interaction can suppress the decoherence effect and bring the state
of nucleotide base to a superposition of all tautomer forms. Indeed,
superposition of all tautomer forms is nothing else then intrabase
entanglement of the atoms on $WC$ edge.

Base pairing occurs via hydrogen bonding, but nature of the
interbase hydrogen bonds is not well understood because of the
insensitivity of experiments. A nonlocal DFT method \cite{HBK-1}
showed that covalent contribution to hydrogen bonds is 38\% in
A$\cdot$T pairing and is 35\% in G$\cdot$C pairing. This conclusion
was then supported by subsequent DFT studies \cite{HBK-7} and
similar conclusions were reached by semiemprical methods with
geometrical and atoms-in-molecules topological parameters, natural
bond orbital analysis, and spectroscopic measurements \cite{HBK-5}.
Covalency of hydrogen bonds means that electron lone pair of the
acceptor orbital is quantum mechanically shared between its own
orbital and unoccupied antibonding orbital linking donor and
hydrogen atoms. Thus, we can interpret the state of a hydrogen
bonded atom pair as an entangled state. Hence, in our model, bases
are paired by swapping intrabase entanglements with interbase
entanglements.

\begin{figure}[h]
\centering
\includegraphics[width=0.48\textwidth]{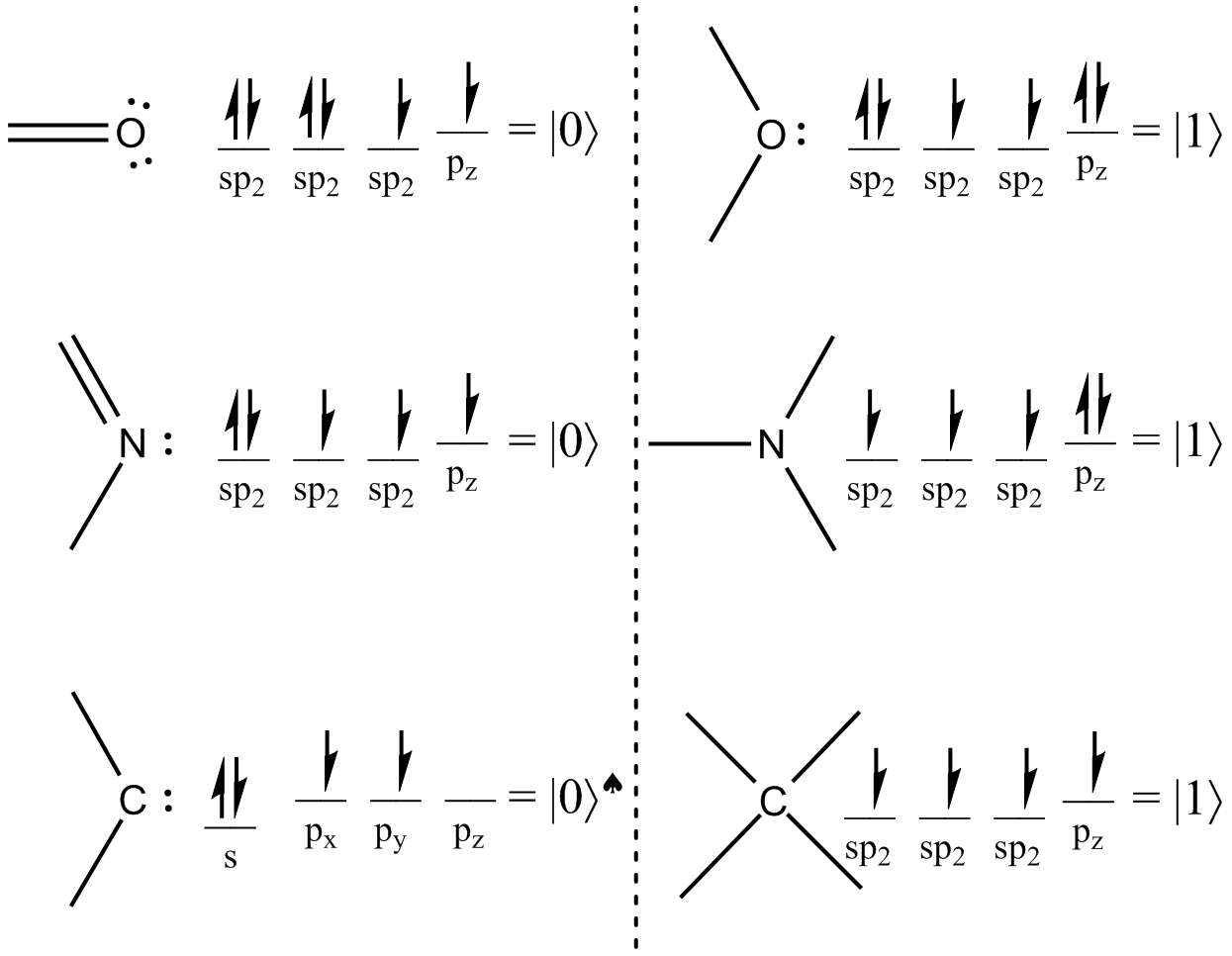}
\caption{Electronic configurations and qubit representations of the
O, N, and C atoms: configuration indicated by $^{\spadesuit}$ is not
present in any tautomer form. However, it is possible to observe it
in blue-shifting hydrogen bonds of DNA.} \label{F-EC}
\end{figure}

We assume that base recognition by DNA\emph{pol} should occur over
the $H$ edges of the nucleotides via a quantum measurement. In
consensus, hydrogen bond donor and acceptor atoms of bases are only
O and N atoms. However, there are a small number of computational
observations in which C atoms of nucleotide bases have the ability
to make blue-shifting hydrogen bonds \cite{HBPT-3}. In this respect,
when electronic configurations of the individual O, N, and C atoms
in $H$, $WC$, and $S$ edges are considered, it is found that each
atom has two different energy states: a relatively lower energy
state for acceptor situation and a relatively higher energy state
for donor situation (Figure \ref{F-EC}). Surprisingly, only $H$ of
the nucleotides can be represented as four orthogonal states if
lower energy states are regarded as $|0\rangle$ qubit and higher
energy states are regarded as $|1\rangle$ qubit:
\begin{eqnarray}
|A\rangle_{H} \!\! &=& \!\! |01\rangle \, , \qquad |T\rangle_{H}
=|10\rangle \, ,
\\
|G\rangle_{H} \!\! &=& \!\! |00\rangle \, , \qquad |C\rangle_{H} =
|11\rangle \, . \nonumber
\end{eqnarray}

The first qubit carries information about purine-pyrimidine
distinction, whereas the second one carries information about
imino-enol distinction. In this sense, DNA\emph{pol} should pair
bases whose qubit representations are complementary to each other.
Not only correct base pairings, but also mispairings like
A$\cdot$C$^{*}$ and G$^{*}$$\cdot$T pairings can be accounted for by
this assumption.

However, since pairing occurs between $WC$ edges of nucleotides,
information processing starting with recognition over $H$ edges
should continue over the $WC$ edges. Equality of the second qubit of
$H$ edge and the first qubit of $WC$ edge makes such a transition in
interaction region reasonable. Qubit representation of $WC$ edges of
nucleotides before replication are as follows:
\begin{eqnarray}
\label{E-I} |A\rangle_{WC,I} \!\! &=& \!\! |101\rangle \, , \qquad
|T\rangle_{WC,I} = |010\rangle \, ,
\\
|G\rangle_{WC,I} \!\! &=& \!\! |011\rangle \, , \qquad
|C\rangle_{WC,I} = |100\rangle \, . \nonumber
\end{eqnarray}

\begin{figure}[h]
\centering
\includegraphics[width=0.40\textwidth]{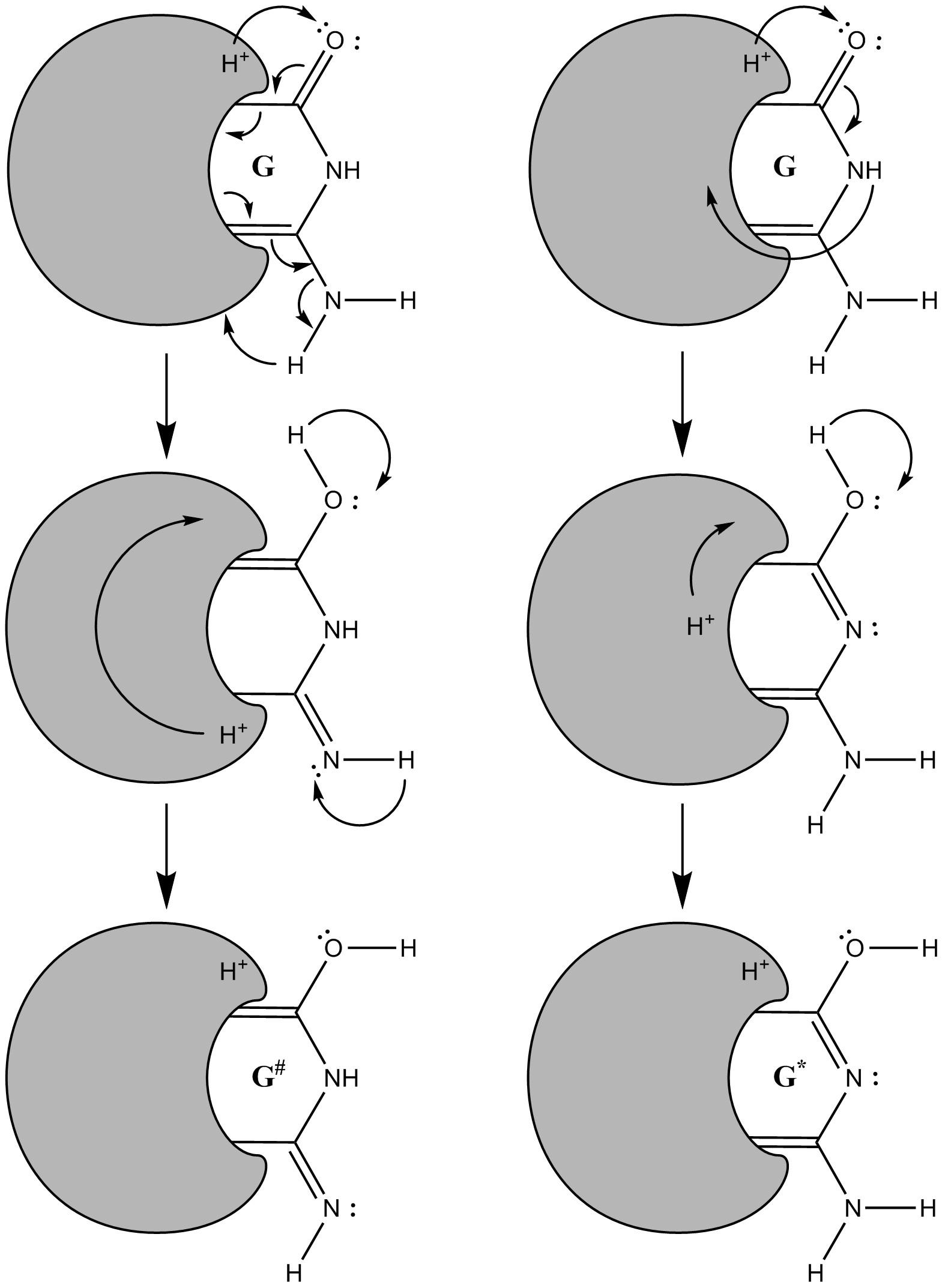}
\caption{Tautomeric transition by proton transfer between enzyme and
nucleotide} \label{F-E}
\end{figure}

\begin{figure}[h]
\centering
\includegraphics[width=0.30\textwidth]{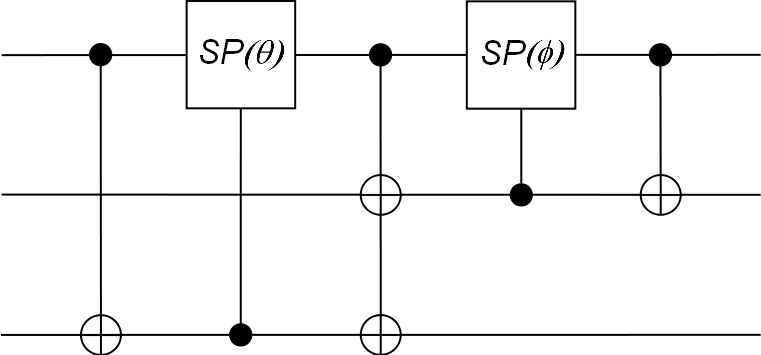}
\caption{A possible quantum circuit for $\mathbf{U}$ which
transforms $|N\rangle_{WC,I}$ states to the $|N\rangle_{WC,Q}$
states: superposition matrix $SP(\theta)$ of
$controlled-Superposition$ gates equals to the multiplication of
rotation matrix $R(\theta)$ and Pauli-$Z$ matrix} \label{F-QC}
\end{figure}

A double proton transfer between the DNA\emph{pol} and the first
atom of $WC$ edge which occurs immediately after the recognition,
can trigger a tautomeric transition (Figure \ref{F-E}). If such a
transfer has a quantum nature, recognition interaction can trigger
an unitary transition to the superposition of all tautomer forms. A
candidate for such a transformation $\mathbf{U}$ is shown in the
Figure \ref{F-QC}. Since $|0\rangle$ and $|1\rangle$ states of an
atom respectively correspond to the absence and presence of a proton
bonded to that atom, \emph{NOT} gate can be regarded as a quantum
mechanical proton transfer. Due to the same reason, \emph{SP} gates
can be considered as formation of a quantum mechanical hydrogen bond
between the enzyme and the particular atom. Both the proton transfer
and hydrogen bonding are the usual tasks done by enzymes and there
are some evidences for the unignorable role of quantum effects and
dynamics on the enzymatic reactions \cite{KE-11}. This means that
intrabase entanglement by transformation $\mathbf{U}$ is a possible
action for the enzyme DNA\emph{pol}.

To provide an equilibrium between maximal entanglement and
robustness, we take the angles $\theta$ and $\phi$ in the quantum
circuit for $\mathbf{U}$ as $\arccos(\sqrt{2} / \sqrt{3})$ and
$\arccos(1 / \sqrt{2})$, respectively. Then, $|N\rangle_{WC,Q}$
states are obtained as:
\begin{eqnarray} \label{E-1}
|A\rangle_{WC,Q} &=& (|01\rangle-|10\rangle)|1\rangle/\sqrt{2} \, ,
\\
|T\rangle_{WC,Q} &=& (|01\rangle+|10\rangle)|0\rangle/\sqrt{2} \, ,
\nonumber
\\
|G\rangle_{WC,Q} &=& (|011\rangle+|101\rangle+|110\rangle)/\sqrt{3}
\, , \nonumber
\\
|C\rangle_{WC,Q} &=& (|100\rangle-|010\rangle+|001\rangle)/\sqrt{3}
\, . \nonumber
\end{eqnarray}

To consider each base pair as an intact system, tensor products of
these states should be taken. However, we reorder qubits of these
product states in such a way that hydrogen bonded atom pairs come
next to each other in order to clarify base pairing. Then, we get
\begin{eqnarray} \label{E-3}
|A\cdot T\rangle_{WC,Q} =\frac12
(\!\!\!\!&|00\rangle|11\rangle|10\rangle&\!\!\!\!+|01\rangle|10\rangle|10\rangle
\\ -\!\!\!\!&|10\rangle|01\rangle|10\rangle&\!\!\!\!-|11\rangle|00\rangle|10\rangle) \, , \nonumber
\\
|G\cdot C\rangle_{WC,Q} =\frac13
(\!\!\!\!&|01\rangle|10\rangle|10\rangle&\!\!\!\!+|11\rangle|00\rangle|10\rangle\!+\!|11\rangle|10\rangle|00\rangle
\nonumber
\\ -\!\!\!\!&|00\rangle|11\rangle|10\rangle&\!\!\!\!-|10\rangle|01\rangle|10\rangle\!-\!|10\rangle|11\rangle|00\rangle \nonumber
\\ +\!\!\!\!&|00\rangle|10\rangle|11\rangle&\!\!\!\!+|10\rangle|00\rangle|11\rangle\!+\!|10\rangle|10\rangle|01\rangle) \nonumber
\end{eqnarray}

\begin{figure}[h]
\centering
\includegraphics[width=0.48\textwidth]{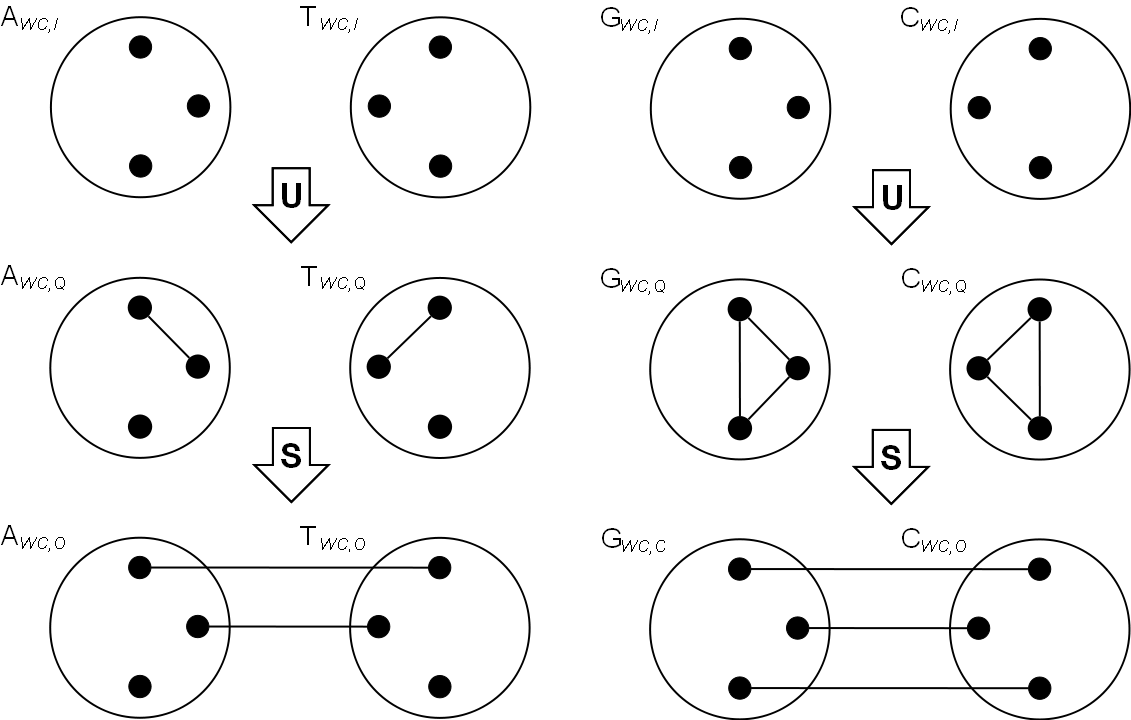}
\caption{Entanglement swapping model of replication} \label{F-M}
\end{figure}

If intrabase hydrogen bonds have a quantum nature as discussed, each
hydrogen bonded atom pair of two paired nucleotides
(N$_1$$\cdot$N$_2$)$_{WC,O}$ can be considered to be in an entangled
state. Then, in order to turn intrabase entanglements into interbase
entanglements, $\mathbf{U}$ should be followed by an irreversible
transformation, $\mathbf{S}:|N_1\cdot
N_2\rangle_{WC,Q}\rightarrow|N_1\cdot N_2\rangle_{WC,O}$, which is
an entanglement swapping (Figure \ref{F-M}). We observe that, in the
case of G$\cdot$C pair, before $\mathbf{S}$ there are two
three-qubit (intrabase) entanglements and after $\mathbf{S}$ there
are three two-qubit (interbase) entanglements. Similarly, in the
case of A$\cdot$T pair, before $\mathbf{S}$ there are two two-qubit
entanglements and after $\mathbf{S}$ there are two two-qubit
entanglements.

\begin{figure}[h]
\centering
\includegraphics[width=0.20\textwidth]{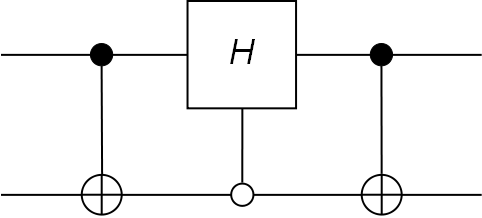}
\caption{Transformation $\mathbf{V}$ in the protocol which swaps
intrabase entanglements to interbase entanglements: this
transformation entangles the qubits on the condition of their
equality. $Hadamard$ matrix $H$ equals to $SP(\pi/4)$.}
\label{F-QC2}
\end{figure}

Swapping intrabase entanglements to interbase entanglements can be
achieved by a five-step protocol $\mathbf{S}$ as follows:

1. Third and fifth qubits of the reordered base pair states (second
and third qubits of the nucleotide base in template DNA) are
subjected to a transformation $\mathbf{V}$ as shown in Figure
\ref{F-QC2}.

2. A Bell measurement is performed on the third and fourth qubits of
the reordered states.

3. If the result of the measurement is one of the two Bell states
$|\beta_{00}\rangle$ and $|\beta_{10}\rangle$ ($(|00\rangle \pm
|11\rangle)/\sqrt{2}$), fourth and fifth qubits of the reordered
states are subjected to the Pauli-$X$ transformation.

4. A Bell measurement is performed on the first and second qubits of
the reordered states.

5. If the result of the measurement is one of the two Bell states
$|\beta_{00}\rangle$ and $|\beta_{10}\rangle$, second and fifth
qubits of the reordered states are subjected to the Pauli-$X$
transformation.

If possible $|N_1\cdot N_2\rangle_{WC,O}$ states are written in
terms of the Bell states $|\beta_{01}\rangle$ and
$|\beta_{11}\rangle$ ($(|00\rangle \pm |11\rangle)/\sqrt{2}$), then
state ensembles of base pairs after $\mathbf{S}$ are found to be
\begin{eqnarray} \label{E-4}
|A\cdot T\rangle_{WC,O} =
\{0.43\!\!\!&,&\!\!\!|\beta_{01}\rangle|\beta_{01}\rangle|10\rangle
;
\\
0.07\!\!\!&,&\!\!\!|\beta_{11}\rangle|\beta_{01}\rangle|10\rangle ;
\nonumber
\\
0.07\!\!\!&,&\!\!\!|\beta_{01}\rangle|\beta_{11}\rangle|10\rangle ;
0.43,|\beta_{11}\rangle|\beta_{11}\rangle|10\rangle\} \, , \nonumber
\\
|G\cdot C\rangle_{WC,O} =
\{P_{jm}^{l}\!\!\!&,&\!\!\!|\beta_{jk}\rangle|\beta_{mn}\rangle(a_{jm}^{l}|01\rangle
+ b_{jm}^{l}|10\rangle)\} \, . \nonumber
\end{eqnarray}
\begin{table}[h]
\centering \caption{Probabilities and probability amplitudes in
state ensemble of G$\cdot$C pair}
\begin{tabular}[c]{|c||c|c|c||c|c|c|}
\hline $l$ & $a_{00}^{l}$ & $b_{00}^{l}$ & $P_{00}^{l}$ &
$a_{10}^{l}$ & $b_{2,10}^{l}$ & $P_{10}^{l}$
\\
\hline 1 & +0.51 & $-$0.86 & 0.11 & $-$0.51 & +0.86 & 0.11
\\
2 & +0.38 & +0.92 & 0.09 & $-$0.38 & $-$0.92 & 0.09
\\
3 & +0.96 & $-$0.28 & 0.03 & $-$0.96 & +0.28 & 0.03
\\
4 & $-$0.92 & +0.38 & 0.02 & +0.92 & $-$0.38 & 0.02
\\
\hline \hline $l$ & $a_{01}^{l}$ & $b_{01}^{l}$ & $P_{01}^{l}$ &
$a_{11}^{l}$ & $b_{11}^{l}$ & $P_{11}^{l}$
\\
\hline 1 & +0.51 & +0.86 & 0.11 & +0.51 & +0.86 & 0.11
\\
2 & +0.38 & $-$0.92 & 0.09 & +0.38 & $-$0.92 & 0.09
\\
3 & $-$0.96 & $-$0.28 & 0.03 & $-$0.96 & $-$0.28 & 0.03
\\
4 & +0.92 & +0.38 & 0.02 & +0.92 & +0.38 & 0.02
\\
\hline
\end{tabular}
\label{T-GC}
\end{table}

We note that, besides $\mathbf{U}$, $\mathbf{S}$ also consists of
only proton transfer and hydrogen bonding processes, excluding Bell
measurements. Bell measurements can be thought as formation of a
quantum mechanical hydrogen bond between measured atom pair if the
outcome state is $|\beta_{01}\rangle$ or $|\beta_{11}\rangle$.
However, when the state of an atom pair collapses to
$|\beta_{00}\rangle$ or $|\beta_{10}\rangle$, atom pair and
DNA\emph{pol} can not separate from each other since total proton
number of base pair does not remain constant after the measurement.
Thus, Bell measurements should be treated as formation of quantum
mechanical hydrogen bonds between the enzyme and measured atom pair
if the outcome state is $|\beta_{00}\rangle$ or
$|\beta_{10}\rangle$. In such circumstances, conditional Pauli-$X$
transformations can fix the total number of protons on base pair and
make atom pair - enzyme complex separable.

Neither $\mathbf{U}$ nor $\mathbf{S}$ is unique for the given model.
However, this is not a disadvantage since there are several
DNA\emph{pol} species and families with different replication
fidelities. This diversity in replication fidelity of DNA\emph{pol}
can be accomplished by different $\mathbf{U}$ and $\mathbf{S}$
pairs.

Since all of the states in both $\mathbf{U}$ and $\mathbf{S}$ can be
expressed as proton transfer and hydrogen bonding, our model could
be tested by repeating the scenario with quantum chemical
computations. Experimental verification is also possible. Evolution
of the $|N_1\cdot N_2\rangle_{WC,O}$ states in the presence of
double well potentials can be prevented by sufficiently decreasing
the time period between two successive pairings. Then, probability
of point mutations due to the formation of rare tautomer forms may
increase according to Equation \ref{E-4} and Table \ref{T-GC}.

Entanglement swapping may be a basic tool used by enzymes and
proteins in the cellular environment. If so, similar models may be
developed for amino acid -- tRNA, aminoacyl-tRNA -- mRNA, and amino
acid -- amino acid interactions in the protein synthesis. If
successful models for these interactions can be developed, then we
can achieve a deeper understanding of the role of the quantum
effects and dynamics on the cellular information processing.

\begin{acknowledgements}
Onur Pusuluk acknowledges support from T\"{U}B\.{I}TAK National
Scholarship Program for Ph.D. Students.
\end{acknowledgements}

\end{document}